\documentclass[twocolumn,aps,pra]{revtex4}
\usepackage{graphicx}% Include figure files
\usepackage{amsmath}
\usepackage{tipa}
\usepackage{amssymb}
\usepackage{color}

\begin{document}

\title{Multiple-Channel Scattering Resonance of One-Dimensional Ultracold Spinor Bosons}
\author{Xiaoling Cui}
\affiliation{Beijing National Laboratory for Condensed Matter Physics,
Institute of Physics, Chinese Academy of Sciences, Beijing 100190, China}
\date{\small \today}

\begin{abstract}
So far the interaction of ultacold atoms can only be tuned within
one particular scattering channel near a resonance, where the spinor
structure of atomic isotopes is destroyed due to the typically large
magnetic field. In this work, we propose a scheme to realize {\it
multiple-channel  scattering resonance} (MCSR) of ultracold bosons
in one-dimension while still keeping their spinor structure. The
MCSR refers to a simultaneous scattering resonance among all
different scattering channels, including those breaking SU(2) and
SO(2) spin rotation symmetries.  Essential ingredients for MCSR
include the 3D interactions, the confinement potential and a
spin-flipping field. Near MCSR a many-body spinor system exhibits
exotic spin density distributions and pair correlations, which are
significantly different from those near a single-channel resonance.

\end{abstract}

\maketitle

\section{Introduction}
Easy access to strong coupling regime of interacting particles and full liberation of the bosonic and fermionic spin degree of freedom comprise two unique and important features of ultracold atomic gases.
%For the former, the widely used approaches to strong couplings include the techniques of Feshbach resonance(FR) in three-dimension(3D)\cite{Chin} and the confinement-induced-resonance(CIR) in low-D\cite{Olshanii, Petrov, CIR_expe1, CIR_expe2, Kohl}.
Especially, the former allows the exploration of intriguing properties of strongly correlated many-body systems, such as the BCS-BEC crossover, the universal thermodynamics, and the Tonks and super-Tonks continuity of one-dimensional(1D) gas\cite{Giorgini, Bloch, Guan}. For the latter, taking advantage of the high (hyperfine) spin structure of atomic isotopes and the SU(2)-invariant interaction at low fields, the atomic spinor system has been shown to exhibit diverse spin textures in the ground state\cite{Ho, Japan, Ketterle, Ho_spin2, Ueda_spin2}, and interestingly coherent spin-exchange dynamics\cite{Pu, Chapman1, Chapman2, Bloch_spin2, Sengstock_spin2, Sengstock_fermion}.
% the degenerate hyperfine spin structure and the SU(2)-invariant interaction facilitate the observation of rich ground state spin texture and the coherent spin-exchange dynamics in spinor systems

Despite all these achievements, little attention has been paid to spinor system with strong interactions\cite{1d_spinor}.
%, i.e., combining both of the two important factors  of cold atomic gases.
An essential reason is this requires the combination of both accessing to strong coupling regime and keeping the spinor structure unchanged, which are hard to realize simultaneously in the experiments so far. Specifically, the widely used approaches to strong couplings involve the techniques of Feshbach resonance(FR) in 3D\cite{Chin} and the confinement-induced-resonance(CIR) in low-D\cite{Olshanii, Petrov, CIR_expe1, CIR_expe2, Kohl}. Both of them can only tune the interaction in one particular spin-collision channel, but not the others, %in a significant way
near a selected resonance\cite{Chin, CIR_expe1, CIR_expe2, Kohl}.
%these two factors seem to be incompatible. First, the FR and CIR can only tune the interaction significantly in one particular spin-collision channel, but not the others, near a selected resonance\cite{Chin, CIR_expe1, CIR_expe2, Kohl}.
The residue symmetry is thus SO(2) symmetry with only the total magnetization conserved but not the total spin anymore. Even worse, near these resonances the magnetic field is typically as large as hundreds of Gauss, where the system tends to be fully polarized by the large Zeeman splitting and thus loses the spinor structure (which requires Zeeman splitting much smaller than interaction energy\cite{Ho}).

In this work, we aim at generating strong coupling in multiple scattering channels with the spinor structure still maintained. Specifically, we propose a two-species bosonic spinor system in 1D geometry with strong interactions. Here the advantage of 1D geometry is that the atom loss is strongly suppressed at strong couplings\cite{Haller}, in contrary to the 3D counterpart. To realize such a system, we apply a radio-frequency(rf) field and an external magnetic field, which respectively induces spin flips and tunes the interaction in a single scattering channel. We show that this system exhibits new physics incorporating both features of multiple-spin degree of freedom and strong coupling of particles, which manifests themselves in generating exotic low-energy scattering properties and significant many-body effects, as summarized below:

{\bf (A)} The low-energy effective scattering will break SU(2) and SO(2) symmetries, i.e., the scattering process will no longer conserve any component of the total spin of incident particles. %$F^2$ or any $F_{\alpha}(\alpha=x,y,z)$.

{\bf (B)} By tuning the rf field or magnetic field, all scattering channels will simultaneously go across the resonance, named as {\it multiple-channel scattering resonances (MCSR)}.

{\bf (C)} Near MCSR, a many-body system exhibits very different properties from those near a single-channel resonance. First, the spin-flip process is greatly enhanced,
%spin densities tend to be balanced between different species,
even in the presence of a weak rf field. Secondly, despite of the strong intra-species repulsion, the system exhibits evidently attractive correlations.  %and even be possible to turn purely attractive.
These properties are experimentally detectable through the measurements of spin densities and two-body correlation functions.
% to  possible to drive the large off-$S_z$ scattering is possible to drive the system to attractively interacting regime.

The rest of the article is organized as follows. We set up the model Hamiltonian for our system in section II, and present the formula for solving the two-body problem in section III. The result of MCSR and its mechanism based on a two-channel model are discussed in section IV, and its many-body effect is studied in section V. Finally we summarize our results in section VI.

\section{Model}

We consider two-species bosons (denoted as $\uparrow,\ \downarrow$) in 1D geometry subject to tight transverse harmonic traps (with frequency $\omega_{\perp}$) and a rf field (with strength $\Omega$). The Hamiltonian for two such atoms located at (${\bf r_1,\ r_2}$) is given by $H=\sum_{i=1}^2 H^{(0)}_i+U$, where
\begin{eqnarray}
%H&=&\sum_{i=1}^2 H^{(0)}_i+U; \label{h3d}\\
H^{(0)}_i&=&-\frac{\bigtriangledown_{x_i}^2+\bigtriangledown_{y_i}^2}{2m}+\frac{m}{2}\omega_{\perp}^2 (x_i^2+y_i^2)+h_i^{(0)},\label{h3d} \\
U&=&\sum_{M=1,0,-1} U_{MM}\delta({\bf r}_1-{\bf
r}_2)|M\rangle\langle M|.\label{u3d}
\end{eqnarray}
Here
\begin{equation}
h^{(0)}_i=-\frac{\bigtriangledown_{z_i}^2}{2m}+\Omega\sigma_x^i
\end{equation}
is the non-interacting hamiltonian along the 1D tube ($z$), and $\sigma_x$ is Pauli matrix inducing spin-flip.
%, and $H_i^{\text{NAEF}}$ is the single-particle NAEF given by case (a) or (b); the dispersion of $h^{(0)}$ is given in Fig.1, which generally have two branches, the lower $(-)$ and the upper $(+)$ one.  %given by $\Omega\sigma_x$ if generated from a rf field and $qp_z\sigma_z+\Omega\sigma_x$ is from Raman-induced spin-orbit coupling. The energy spectrum for each $H_{NAF}$ is shown in Fig.1;
$U$ characterizes scattering in three channels classified by
 total magnetization $M$, and specifically
\begin{eqnarray}
|M=1\rangle&=&|\uparrow_1\uparrow_2\rangle, \\
|M=0\rangle&=&\frac{1}{\sqrt{2}} (|\uparrow_1\downarrow_2\rangle+|\downarrow_1\uparrow_2\rangle),\\
|M=-1\rangle&=&|\downarrow_1\downarrow_2\rangle.
\end{eqnarray}
$U$ in each $M$-channel is associated with a s-wave scattering length $a_{M}$,
via
\begin{equation}
\frac{1}{U_{MM}}=\frac{4\pi a_{M}}{m}-\frac{1}{V} \sum_{\bf k} \frac{m}{k^2},
\end{equation}
where $V$ is the volume.

To realize above model Hamiltonian, one can use $F=1$ alkali isotopes such as ${}^{87}$Rb, with $|m_F=1(0)\rangle\equiv |\uparrow(\downarrow)\rangle$.  Through a FR at $B_0=1007G$, the scattering length $a_1$ can be tuned efficiently (but not $a_0,\ a_{-1}$). At magnetic field $B\sim B_0$, the large Zeeman splitting between $\uparrow$ and $\downarrow$ spins can be effectively eliminated by applying a rf field and tuning its frequency on resonance with the splitting. In the frame rotating at the rf frequency, a spinor bosonic system (without Zeeman splitting due to external $B$) can be created.
Similar setup has been realized in a previous experiment\cite{Peierls}.
Note that the third component of $F=1$ isotopes, $|m_F=-1\rangle$, can be adiabatically eliminated due to the large quadratic Zeeman shift at $B\sim B_0$\cite{footnote_Rb}.

For the low-energy scattering, we will show that the system can be described by an effective 1D Hamiltonian: $h=\sum_{i=1}^2 h^{(0)}_i+u$, with
\begin{eqnarray}
u&=&\sum_{MN} u_{MN}\delta(z_1-z_2)|M\rangle\langle N|.\label{h1d}
\end{eqnarray}
Here $u_{MN}$ is the effective 1D coupling strength between channel $M$ and $N$, and we have expressed Eq.(\ref{h1d}) in a most general form with off-magnetization scattering terms (i.e., $u_{MN}$ with $M\neq N$). As we will show in this work, these terms do exist in $u$, hence demonstrating a unique scattering property of the confined 1D system in comparison with the 3D one (see $U$ in Eq.(\ref{u3d})). In the following we will calculate $u$ by matching the two-body solutions of $H$ and $h$, with the only criterion that they produce the same low-energy scattering property.

\section{Two-body formulism}

We study the full scattering wavefunction $|\Psi\rangle$ according to  $H|\Psi\rangle=E|\Psi\rangle$, with low energy $E\ll E_{th}+2\omega_{\perp}$ where $E_{th}=\omega_{\perp}-2\Omega$ is the threshold energy. We only consider the relative motion here since it is interaction-relevant and can be decoupled from the center-of-mass motion. Given $U$ in (\ref{u3d}), we assume
%Based on T-matrix approach\cite{t-matrix}, we write
\begin{eqnarray}
\langle{\bf r}|U |\Psi\rangle=\delta({\bf r}) \sum_{M=-1,0,1}
F_{M}|M\rangle,\ \ \ \ {\bf r}\equiv {\bf r}_1-{\bf r}_2 \label{U3d}
\end{eqnarray}
Further utilizing the Lippman-Schwinger equation
\begin{equation}
U|\Psi\rangle=T|\Psi^{(0)}\rangle,
\end{equation}
where the T-matrix follows
\begin{equation}
T=U+UG_{0}T,
\end{equation}
$|\Psi^{(0)}\rangle$ is the incident wave function and $G_0=(E-H^{(0)}_1-H^{(0)}_2+i0^+)^{-1}$ is the non-interacting Green function, we obtain the scattering amplitudes $\{F_{M}\}$ in (\ref{U3d}) via following matrix equation\cite{t-matrix}
\begin{eqnarray}
&&\sum_{N} \left[U^{-1}-\tilde{G}_0 \right] _{MN}  F_{N} = \tilde\Psi^{(0)}_M ; \label{f3d}
%&&\tilde{G}_0=\int dZ' e^{-iK(Z-Z')} \langle {\bf 0}; Z |G_0| {\bf 0}; Z' \rangle, \nonumber
\end{eqnarray}
with $\tilde{G}_0=G_0({\bf r=0})$, $\tilde\Psi^{(0)}_M=\langle M|\Psi^{(0)}({\bf r=0}) \rangle$.
Here $U$ and $\tilde{G}_0$ are both $3\times 3$ matrixes expanded in
the $\{|M\rangle\langle N|\}$ spin basis.
%, and particularly $U_{\alpha\beta}=U_{\alpha}\delta_{\alpha\beta}$ is diagonal matrix.
After solving $\{F_{M}\}$ from (\ref{f3d}), we obtain the wave function as
\begin{eqnarray}
%\langle{\bf r};Z |\Psi\rangle=\langle{\bf r};Z |\Psi^{(0)}\rangle+ \sum_{M} F_{M} \int dZ' e^{iKZ'}\langle{\bf r};Z | G_0 |{\bf 0};Z' \rangle |M\rangle.
\Psi ({\bf r})=\Psi^{(0)} ({\bf r})+ \sum_{M} F_{M} G_0 ({\bf r}) |M\rangle.  \label{psi3d}
\end{eqnarray}
When $z\rightarrow\infty$, the wave function (\ref{psi3d}) is frozen at the lowest transverse mode ($n_x=n_y=0$), i.e.,
\begin{equation}
\Psi ({\bf r}) \rightarrow \phi_0(x) \phi_0(y) \psi (z).   \label{relation}
\end{equation}
%where %$\phi_0(x)$ is the ground state of a 1d harmonic oscillator and
Here $\psi$ describes the effective scattering process along $z$,
which can be equally obtained based on the reduced 1D Hamiltonian
$h$. Similarly we define the 1D scattering amplitudes $f_{M}$, and
find that Eqs.(\ref{U3d},\ref{f3d}) are still applicable as long as
$\{ {\bf r},\ \Psi,\ \Psi^{(0)},\ F_{M},\ U,\ G_0\}$ are
respectively replaced by $\{ z,\ \psi,\ \psi^{(0)},\ f_{M},\ u,\
g_0\}$, where $g_0=(E-E_{th}-h^{(0)}_1-h^{(0)}_2+i0^+)^{-1}$ is
non-interacting Green function in 1D. Given Eq.(\ref{relation}), we
get  $f_{M}=F_{M}{\phi_0^*}^2(0)$ and finally relate $u$ in
(\ref{h1d}) to $U$ in (\ref{u3d}) as
%the reduced interaction matrix
\begin{eqnarray}
u=|\phi_0(0)|^4 \big[ U^{-1}- \tilde{G}_0^{ex}
\big]^{-1}, \ \ \tilde{G}_0^{ex}=\tilde{G}_0-|\phi_0(0)|^4 \tilde{g}_0. \label{u-mtx}
\end{eqnarray}
Here $\tilde{G}_0^{ex}$ is the
Green function constructed by all excited transverse modes $n_x+n_y>0$. As we will see later, its
structure is essential to induce the multiple-channel scattering resonances in $u$. In Appendix A, we present more details for the derivation and evaluation of matrix equations (\ref{f3d}) and (\ref{u-mtx}).

\begin{figure}
\includegraphics[height=3.2cm,width=8.8cm]{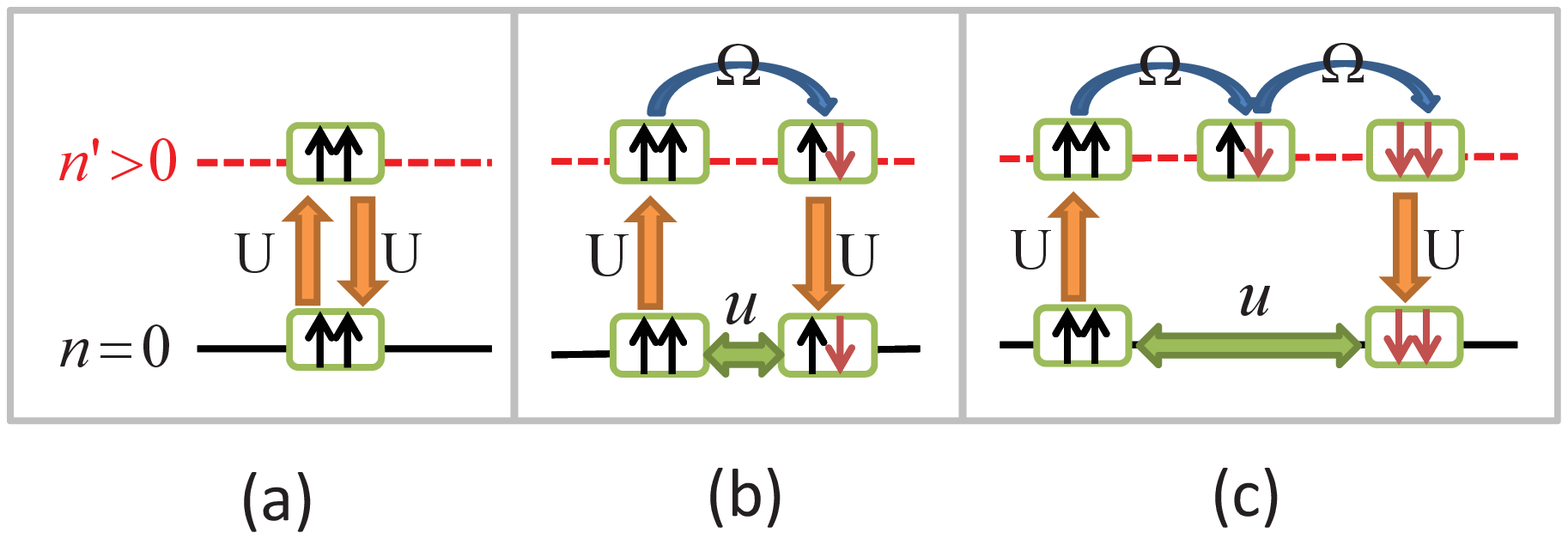}
\caption{(Color Online). Schematic plots of virtual scattering
processes involving higher excited modes without(a) or with(b,c) rf
field. $u_{11}$, $u_{10}$ and $u_{1,-1}$ are respectively
renormalized through the processes in (a), (b) and (c). }
\label{fig1}
\end{figure}

\section{Multiple-channel scattering resonance (MCSR)}

In this section, we present the result of multiple-channel scattering resonance for the reduced 1D coupling strengths, and analyze its physical mechanism based on a two-channel model, which provides a useful estimation on the resonance widths in different scattering channels.

\subsection{Results}

Cooperating with the confinement and 3D interactions, the rf field can result in multi-channel effective scattering in the low-energy 1D space. This is achieved through the virtual scattering processes to higher transverse modes, as schematically shown in Fig.1.
Without rf field (Fig.1a), an initial spin state, $|\uparrow\uparrow\rangle(M=1)$,  at the ground state mode (${\bf n}: \ n_x=n_y=0$),  can only be scattered by 3D interaction $U$ to the same spin state in higher modes (${\bf n'}: \ n_x+n_y>0$), and then back to itself in ${\bf n}$, which process renormalizes the effective $u_{11}$.  When rf field is switched on, two additional processes can occur (Fig.1b and 1c). The rf field could flip spins in {\bf n'} once or twice and finally be scattered to a different spin state $|\uparrow\downarrow\rangle(M=0)$ or $|\downarrow\downarrow\rangle(M=-1)$ in the ground mode ${\bf n}$. These processes respectively renormalize $u_{10}$ and $u_{1,-1}$, which are originally absent if without rf field.
%This process gives rise to an effective off-$F_z$ scattering in the effective 1D scattering.
%This minimal picture, resembling the second-order perturbation process, shows physically how the interplay of rf field, 3D interaction and the confinement  renormalize the coupling strengths and result in multi-channel effective scattering in the low-energy space.

More accurately, the physics illustrated above can be reflected in the exact expression of $\tilde{G}_0^{ex}$(Eq.(\ref{u-mtx})), which include both diagonal and off-diagonal elements contributed from all orders of scattering processes involving all excited modes.  Consequently, $u-$matrix also have non-zero off-diagonal elements, which break both SU(2) and SO(2) symmetries in the spin-spin scattering process. This exactly demonstrates the multiple-channel scattering as summarized previously by (A) in the introduction.

Moreover, due to the intrinsic entanglement between different scattering processes, Eq.(\ref{u-mtx}) further predicts an exotic phenomenon in the low-energy scattering, namely the  {\it multiple-channel scattering resonances} (MCSR), which refers to a simultaneous divergence of effective couplings in different scattering channels (different $u-$matrix elements). For threshold scattering ($E=E_{th}$), the MCSR occurs when
\begin{equation}
|U^{-1}- \tilde{G}_0^{ex}(E=E_{th},\Omega_{res})|=0.  \label{res}
\end{equation}
Here $\Omega_{res}$ is the strength of rf field required by MCSR. Remarkably, it means that by tuning one single parameter ($\Omega$ or scattering length $a_M$ in an arbitrary $M$-channel), the spinor system can be driven to strongly coupling regime in multiple scattering channels. This demonstrates (B) in the introduction. %Note that compared to the previous studies of coupled-channel scattering\cite{CC} and multi-channel quantum defect theory\cite{QDF} in atomic physics, here the MCSR have substantially different background settings, mechanisms, and physical consequences.
Note that the MSCR here should be distinguished from the coupled-channel scattering\cite{CC} and the multi-channel quantum defect theory\cite{QDF} studied in literature\cite{footnote_dif}.

In Fig.2, we show the general features of MCSR by numerically solving Eqs.(\ref{u-mtx},\ref{res}).  Fig.2(a) gives $\Omega_{res}$ as a function of one interaction parameter $a_{\perp}/a_1$ ($a_{\perp}=\sqrt{2/(m\omega_{\perp})}$ is confinement length), while $a_0$ and $a_{-1}$ are both fixed and far off resonance. This resembles the actual case of $^{87}$Rb in realistic experiments\cite{footnote_Rb}. At $\Omega=0$, different scattering channels are decoupled and we recover the result of CIR within the single $M=1$ channel at $a_{\perp}/a_1=1.46$\cite{Olshanii}. At finite $\Omega$, it is found that the resonance position shifts to BEC side with larger $a_{\perp}/a_1$. More importantly, the structure of $u$ is drastically different from what CIR predicted (dashed curves in Fig.2(b1,b2,c1,c2)).  Especially, by tuning $\Omega$ we find non-zero off-$M$ scattering $u_{10}$ ($M=1\leftrightarrow M=0$,  Fig.2(b2)), with its strength approaching resonance regime simultaneously with $u_{11}$ ($M=1\leftrightarrow M=1$, Fig.2(b1)). These multiple resonances can also be achieved by tuning $a_{\perp}/a_1$ while keeping $\Omega$ fixed (Fig.2(c1,c2)).

%fig.2:
\begin{figure}
\includegraphics[height=5cm,width=8.8cm]{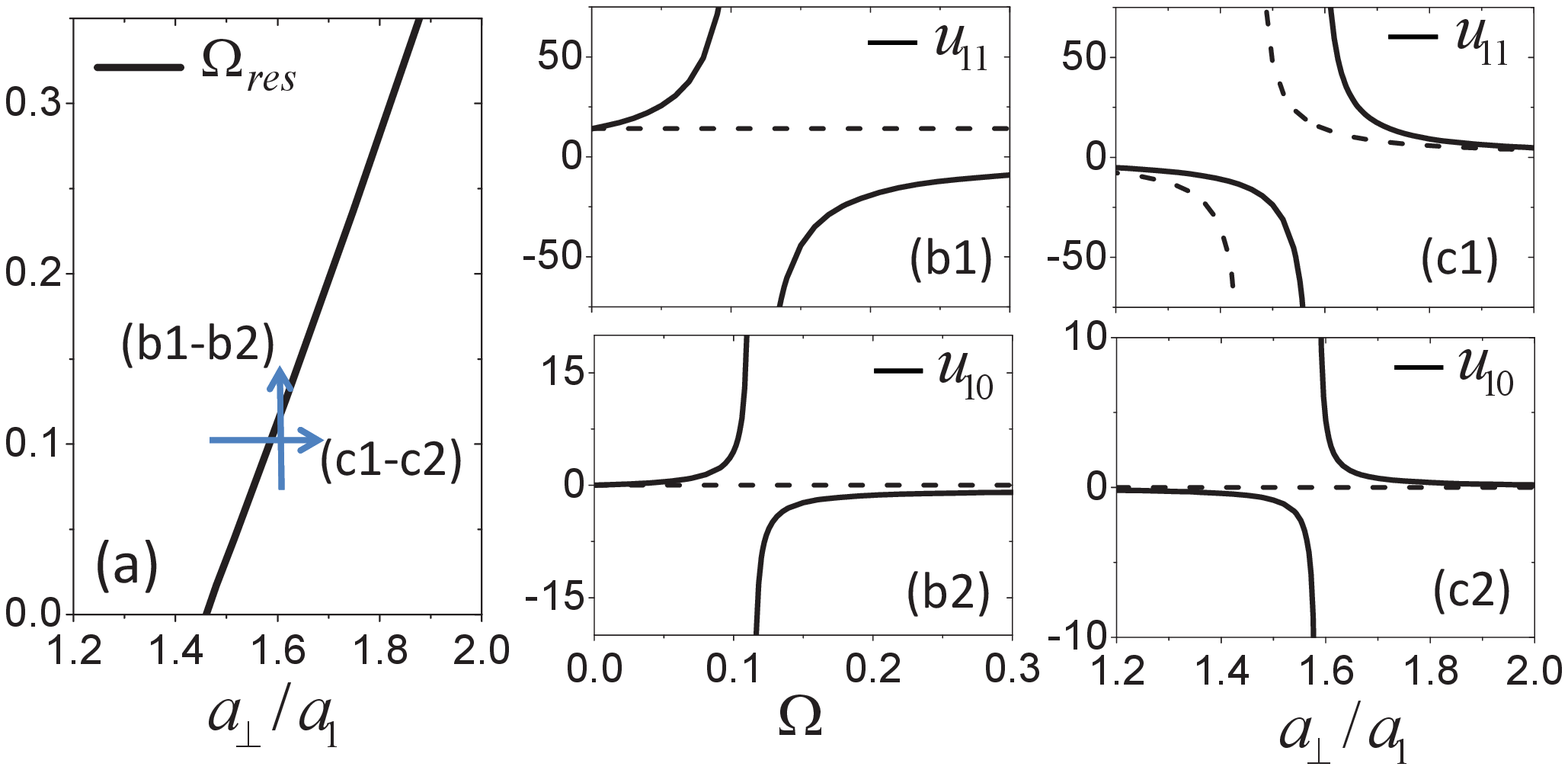}
\caption{(Color Online). Multiple-channel scattering resonances with tunable $a_1$ and fixed $a_0=a_{-1}=a_{\perp}/4$\cite{footnote_Rb}. $\Omega$ is scaled by $\omega_{\perp}$, and $u_{11},\ u_{10}$  are scaled by $2/(m\omega_{\perp})$.
(a): Resonance position $\Omega_{res}$ as functions of $a_{\perp}/a_1$. (b1,b2) [or (c1,c2)]: $u_{11},\ u_{10}$  as functions of $\Omega$ at fixed $a_{\perp}/a_1=1.6$ [or as functions of $a_{\perp}/a_1$ at fixed $\Omega=0.1\omega_{\perp}$], corresponding to the blue vertical [or horizontal] arrow in (a).  For comparison, CIR predictions\cite{Olshanii} are shown by dashed lines. } \label{fig2}
\end{figure}

\subsection{Physical mechanism of MCSR}

Following the traditional way in understanding FR\cite{KK} and CIR\cite{Olshanii}, a thorough physical interpretation for MCSR can also be obtained through a two-channel model, where an open channel and a closed channel are introduced respectively with projection operators $P$ and $Q$. Here the open ($P$) channel refers to scattering within the lowest transverse mode ($n_x=n_y=0$), while the closed ($Q$) channel refers to scattering involving higher transverse modes ($n_x+n_y>0$). With projections $P$ and $Q$, the two-body schrodinger equation can be divided into two equations,
\begin{eqnarray}
H_{PP}\Psi_P+H_{PQ}\Psi_Q&=&E\Psi_P;\\
H_{QP}\Psi_P+H_{QQ}\Psi_Q&=&E\Psi_Q,
\end{eqnarray}
with $H_{\mu\nu}=\mu H\nu$ and $\Psi_{\mu}=\mu\Psi$ ($\mu,\nu=P$ or $Q$). By solving these equations, one can obtain the effective schrodinger equation for the open-channel state $\Psi_P$ as $H_{eff}\Psi_P=E\Psi_P$, with effective Hamiltonian
 \begin{eqnarray}
H_{eff}=H_{PP}+ H_{PQ} \frac{1}{E-H_{QQ}} H_{QP}. \label{Heff}
\end{eqnarray}
The second term of above equation incorporates all contributions from the virtual scattering processes involving higher transverse modes (closed channel), which renormalize the effective scattering within the lowest transverse mode (open channel). As the eigen-value of $H_{QQ}$ can be adjusted by interaction parameters or the strength of rf field, it can be tuned  crosses $E_{th}$ and cause a divergence of $H_{eff}$ according to Eq.(\ref{Heff}). This gives rise to the scattering resonance in the open channel.

\begin{figure}
\begin{center}
\includegraphics[height=4cm,width=8.5cm]{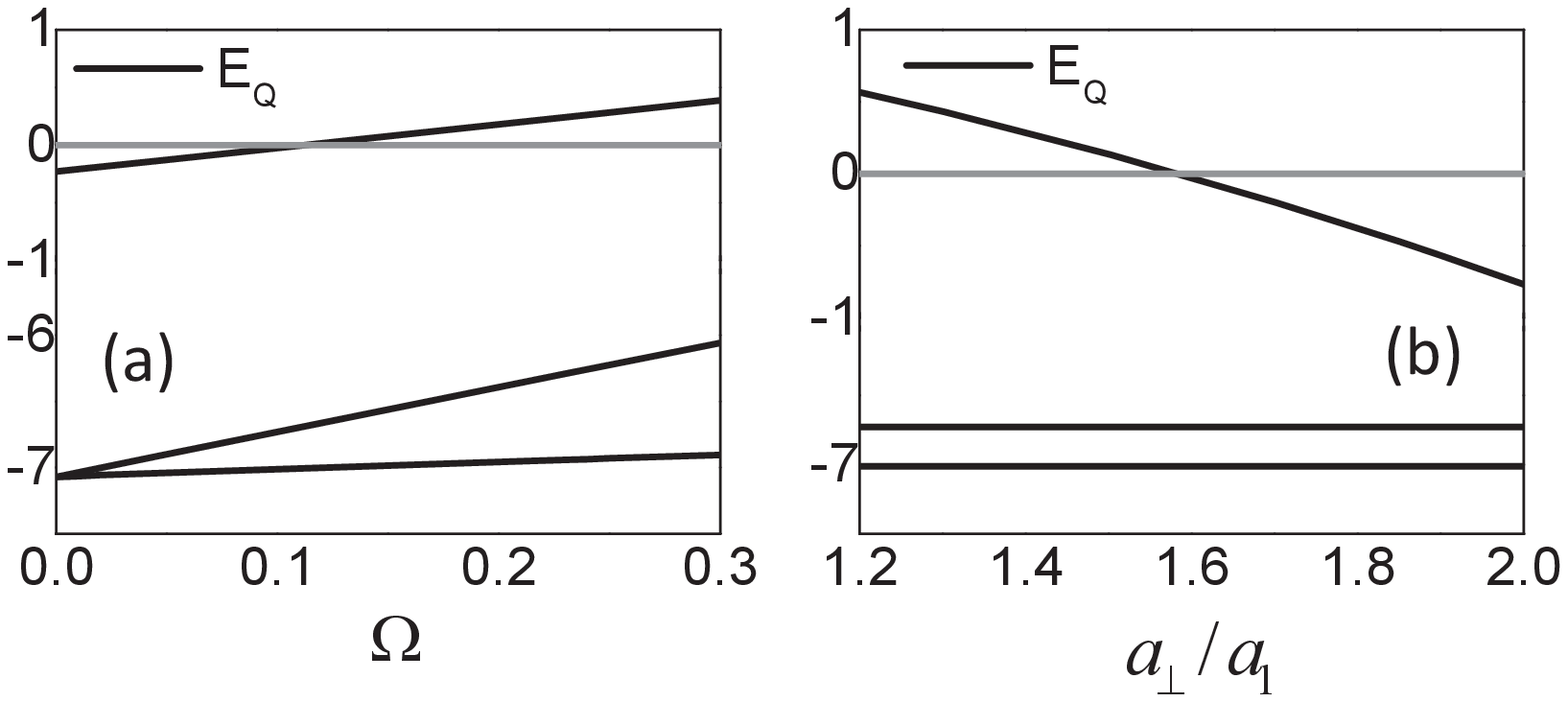}
 \caption{Virtual bound state energies $E_Q$(shifted by $E_{th}$) as functions of $\Omega$ at fixed $a_{\perp}/a_1=1.6$ (a), or as functions of $a_{\perp}/a_1$ at fixed $\Omega=0.1\omega_{\perp}$ (b). Here (a) and (b) respectively follow the vertical or horizontal arrows in Fig.2(a).  $E_Q,\ \Omega$ are all scaled by $\omega_{\perp}$. }\label{fig_supple_Eb}
\end{center}
\end{figure}

We denote the eigen-states of $H_{QQ}$ as $\tilde{\Psi}_Q$, and write $H_{QQ} \tilde{\Psi}_Q=E_Q \tilde{\Psi}_Q$. $E_Q=E_{th}$ determines the scattering resonances for open channel. In Fig.\ref{fig_supple_Eb}, we plot three $E_Q$ evolving with the rf strength or interaction parameters, and the place when one of these bound states across $E_{th}$ gives the location of MCSR. Different from the closed-channel bound state in a single-channel resonance, here in MCSR each bound state is highly entangled in spin space, which is a certain superposition of all $M-$states.
Whenever such a dressed-spin state across threshold, resonances will simultaneously occur in multiple spin-collision channels, as seen from Fig.2(b1,b2,c1,c2).

Using the formula in Eq.(\ref{Heff}), one can evaluate the resonance width in different collision channels, which is proportional to $H_{PQ}H_{PQ}=\langle \Psi_P |H| \tilde{\Psi}_Q \rangle \langle \tilde{\Psi}_Q |H| \Psi_P \rangle$. Explicitly, in our case we write the $a_1$-tuned resonances as
\begin{eqnarray}
u_{MN}&=&\frac{W_{MN}}{a_{\perp}/a_1-C}, \\
W_{MN}  &\propto& \langle \Psi^M_P |H| \tilde{\Psi}_Q \rangle \langle \tilde{\Psi}_Q |H| \Psi^N_P \rangle, \label{Width}
\end{eqnarray}
where $C$ the resonance position of $a_{\perp}/a_1$, $| \Psi^N_P \rangle$ is the open-channel wave function when projected to $|N\rangle$ spin state, and $W_{MN}$ the resonance width of $u_{MN}$.

Given that $a_1$ can be tuned large through FR, while $a_0$ and $a_{-1}$ are far off
resonances (small positive values), in the vicinity of above 1D
resonances the closed channel is mainly composed by $|M=1\rangle$
states, and its wave function can be estimated through the
perturbation theory, i.e.,
\begin{eqnarray}
\tilde{\Psi}_Q ({\bf r})& = &\tilde{\Psi}_{1}^{(0)} ({\bf r}) |M=1\rangle + \tilde{\Psi}_{0}^{(0)} ({\bf r}) \frac{\sqrt{2}\Omega\langle  \tilde{\Psi}_{0}^{(0)}| \tilde{\Psi}_{1}^{(0)}\rangle}{E_{1}^{(0)}-E_{0}^{(0)}} |M=0\rangle \nonumber\\
&&+  \tilde{\Psi}_{-1}^{(0)} ({\bf r}) \frac{2\Omega^2 \langle  \tilde{\Psi}_{-1}^{(0)}| \tilde{\Psi}_{0}^{(0)}\rangle  \langle  \tilde{\Psi}_{0}^{(0)}| \tilde{\Psi}_{1}^{(0)}\rangle}{(E_{1}^{(0)}-E_{0}^{(0)})(E_{1}^{(0)}-E_{-1}^{(0)}) } |M=-1\rangle , \label{Psi_Q}
\end{eqnarray}
here $\tilde{\Psi}_{M}^{(0)} ({\bf r})$ and $E_{M}^{(0)}$ are respectively the eigen-function and eigen-energy of $H_{QQ}$ in $M\leftrightarrow M$ scattering channel at $\Omega=0$. Given above interaction parameters, we have $E_1^{(0)}\gg E_0^{(0)}=E_{-1}^{(0)}\approx -1/(ma_{0}^2)$ or $\approx -1/(ma_{-1}^2)$. Above perturbation theory is valid when
\begin{equation}
\frac{\Omega}{|E_{1}^{(0)}-E_{0(-1)}^{(0)}|}\ll 1. \label{condition}
\end{equation}

Given Eq.(\ref{Width}) and Eq.(\ref{Psi_Q}), one can obtain the relative widths of  all $\{W_{MN}\}$. For instance, we have the ratios
\begin{eqnarray}
\frac{W_{10}}{W_{11}}&\sim& \frac{\Omega}{E_{1}^{(0)}-E_{0}^{(0)}}; \\
\frac{W_{00}}{W_{11}}&\sim&
\frac{\Omega^2}{(E_{1}^{(0)}-E_{0}^{(0)})^2}; \\
\frac{W_{1,-1}}{W_{11}}&\sim&
\frac{\Omega^2}{(E_{1}^{(0)}-E_{0}^{(0)})(E_{1}^{(0)}-E_{-1}^{(0)})}, \label{Width2}
\end{eqnarray}
From this estimation, one can see that under the condition (\ref{condition}), only $W_{11}$
and $W_{10}$ would have visible widths, while the others (in
comparison to $W_{11}$ and $W_{10}$) are too narrow to be resolve in realistic experiments.

\begin{widetext}

\begin{figure}
%\begin{center}
\includegraphics[height=5.5cm,width=17cm]{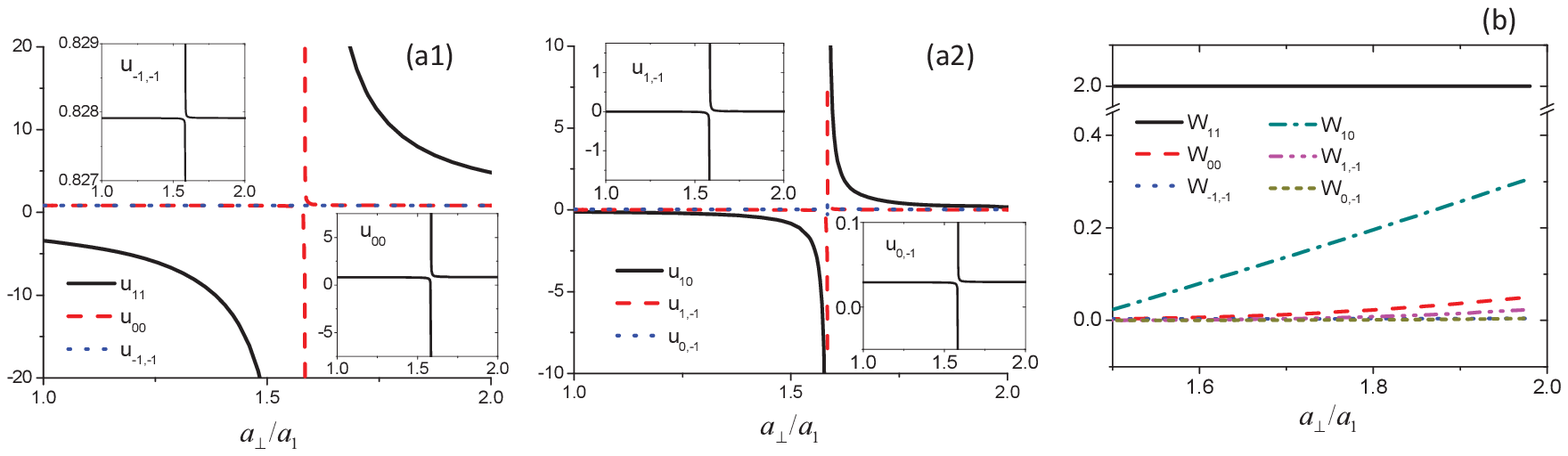}
 \caption{(Color Online). (a1) Diagonal and (a2) off-diagonal matrix elements of $u$ as functions of $a_{\perp}/a_1$ at fixed $\Omega=0.1\omega_{\perp}$. The other parameters,
 $a_0=a_{-1}=a_{\perp}/4$, are the same as those in Fig.2. Insets show magnified plots of $u-$matrix elements except for $u_{11}$ and $u_{10}$. (b) Resonance widths $W_{MN}$ defined in Eq.(\ref{Width}). $u-$elements and $W_{MN}$ are both scaled by $2/(m\omega_{\perp})$.   }\label{fig_supple}
%\end{center}
\end{figure}

\end{widetext}

Above analyses from two-channel model have also been verified by our numerical calculations. Fig.\ref{fig_supple} shows that all the elements of $u$-matrix simultaneously go to infinity as $a_{\perp}/a_1$ across resonance position. However, under the conditions specified before Eq. (\ref{Psi_Q}), only the resonances of $u_{11}$ and $u_{10}$ have visible widths. The other components of $u-$matrix are only large enough if extremely close to the resonance position (with very narrow width).
In the following, we will explore the many-body effect due to strong couplings of $u_{11}$ and $u_{10}$ near MCSR, while the other channels are approximated as non-interacting.

\section{Many-body effects}

Given $u_{11}$ and $u_{10}$ from two-body solutions, we write down the many-body Hamiltonian for 1D spinor bosons subject to an additional harmonic trap,
\begin{eqnarray}
H&=&\sum_{\sigma}\int dz \Psi_{\sigma}^{\dag}(z)\left( -\frac{\hbar^2}{2m}\frac{\partial^2}{\partial z^2} + \frac{1}{2}m\omega_{T} z^2\right) \Psi_{\sigma}(z)  \nonumber\\
&&+ \Omega \int dz \left(\Psi_{\uparrow}^{\dag}(z) \Psi_{\downarrow}(z) + h.c. \right) \nonumber\\
&&+ \frac{u_{11}}{2} \int dz \Psi_{\uparrow}^{\dag}(z)\Psi_{\uparrow}^{\dag}(z)\Psi_{\uparrow}(z)\Psi_{\uparrow}(z)\nonumber\\
&&+\frac{u_{10}}{\sqrt{2}} \int dz \left(\Psi_{\uparrow}^{\dag}(z)\Psi_{\uparrow}^{\dag}(z)\Psi_{\uparrow}(z)\Psi_{\downarrow}(z) + h.c. \right)  \label{H}
\end{eqnarray}
%Note that the two coefficients, $1/2$ and $1/sqrt{2}$, in front of $u_{11}$ and $u_{10}$ is to due to the bosonic
It is clear that the interaction part of $H$ includes various terms breaking SU(2) and SO(2) symmetries, such as $\rho\cdot \sigma_z$, $\rho\cdot \sigma_x$, $\sigma_z\cdot \sigma_z$, $\sigma_z\cdot \sigma_x$, where $\rho$ and $\sigma_{x,z}$ are respectively the number density and spin densities.

Based on Hamiltonian (10), we will first use exact diagonalization method to solve the ground state of a four-particle system. The results obtained not only are relevant to the cluster system\cite{Jochim1,Jochim2}, but also serve as a benchmark for a many-body system. We focus on the large repulsion limit of $u_{11}$, where $u_{10}$ can also be tuned large and positive using MCSR. To highlight the significance of $u_{10}$, we compare three different cases: (i) strongly repulsive spin-$\uparrow$ bosons without rf field ($\Omega=0$) and $u_{10}=0$; (ii) $\Omega\neq 0$ but still $u_{10}=0$; (iii) $\Omega\neq 0$ and $u_{10}\neq0$. Among them, case (iii) is what we are most interested in and also the general one near MCSR.

In Fig.5 we show the spin density distributions,
\begin{equation}
\rho_{\sigma}(z)\equiv\langle \Psi^{\dag}_{\sigma} (z) \Psi_{\sigma} (z) \rangle,
\end{equation}
and the two-body correlation functions,
\begin{equation}
g_{\sigma\sigma'}(z,z_0)\equiv \langle \Psi^{\dag}_{\sigma} (z) \Psi^{\dag}_{\sigma'} (z_0) \Psi_{\sigma'} (z_0) \Psi_{\sigma} (z)\rangle
\end{equation}
with $z_0=0$, for $N=4$ system in cases (i,ii,iii). For case (i), the spin-$\uparrow$ bosons are fermionalized, with wave function well approximated by the absolute value of a slater determinate $\psi_{\uparrow}(\{z\})=\sqrt{\frac{1}{N!}} |Det(\phi_i(z_j))|$ (here $\phi_i$ is the eigen-function of 1D harmonic oscillator, with level index $i=0,...,N-1$). Consequently the density is given by $\rho_{\uparrow}(z)=\sum_{i=0}^{N-1}|\phi_i(z)|^2$, and the two-body correlation by $g_{\uparrow\uparrow}(z,z_0)=\sum_{<i,j>} |\phi_i(z)\phi_j(z_0)-\phi_j(z)\phi_i(z_0)|^2$, which all share the same properties of identical fermions (see gray curves in Fig.5(a1,b1)). When turn on a weak rf, it provides a way to avoid strong repulsion between $\uparrow$-spins by flipping them to $\downarrow$, and thus the ground state will be tremendously changed\cite{footnote_rf}. If $u_{10}$ is absent(case (ii)), rf alone will generate a lot more $\downarrow$ than $\uparrow$ (see Fig.5(a2)). The residual $\uparrow$ are there to take advantage of the rf energy, and they still maintain fermion-like correlations, characterized by a dip of $g_{\uparrow\uparrow}$ at $z=z_0=0$ (Fig.5(b2)).

\begin{figure}
\includegraphics[height=10cm,width=7.5cm]{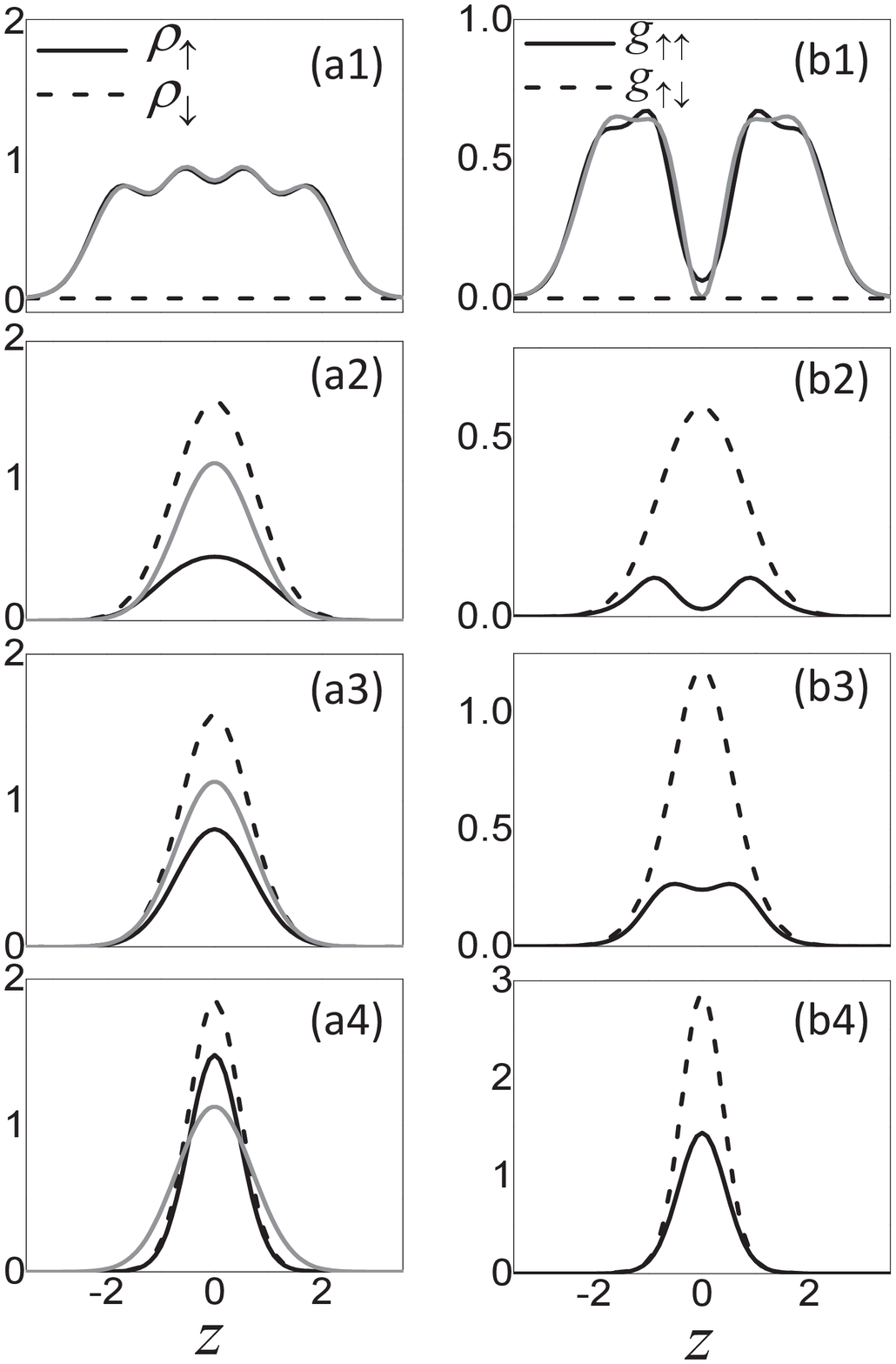}
 \caption{(Color Online). (a1-a4)Spin densities (in units of $1/a_T$, $a_T=(m\omega_{T})^{-1/2}$) and (b1-b4)correlation functions (in units of $1/a_{T}^2$) for four trapped bosons in 1D. $z$ is in unit of $a_T$.
The parameters, $(u_{11}/(a_T\omega_T), u_{10}/(a_T\omega_T), \Omega/\omega_T)$,  are respectively: (a1,b1)$(12,0,0)$; (a2,b2)$(12,0,1)$; (a3,b3)$(12,2\sqrt{2},1)$; (a4,b4)$(12,4\sqrt{2},1)$.
For comparison, gray curves in (a1,b1) show the case of identical fermions(see text), and in (a2-a4) show the density distribution of non-interacting bosons. }\label{fig3}
\end{figure}

Compared with cases (i,ii) which describe a single-channel resonance, the case (iii), with $u_{10}$ present and describing MCSR, will show a very different ground state property (as summarized by (C)). As the $u_{10}$ term follows the form of $\int dz \rho_{\uparrow}(z)\sigma_x(z)$, the spin-flip process will be drastically enhanced through $u_{10}$. As shown in Fig.5(a3) and (a4), the spin numbers get more balanced as $u_{10}$ increases. One may thus expect this $u_{10}$ term just simply enhance the effective rf strength, $\Omega_{\rm {eff}}$. However, this is not true, as it also generates significant interaction effect and modifies the pair correlations. As seen from Fig.5(b3) and (b4), when increasing $u_{10}$ from zero, the original dip of up-spin correlations at $z=z_0=0$ gradually vanishes and turns to a peak, implying the up-spins gradually lose their repulsive nature and turn attractive. Accordingly all spins are attracted to the trap center and produce more pronounced density distribution (than non-interacting case)(Fig.2(a4)). Indeed, assume a many-body system at sufficiently large $u_{10}$, one would expect the spins being polarized along $-\hat{x}$, giving $\sigma_x(z)=-2\rho_{\downarrow}(z)=-2\rho_{\uparrow}(z)$. Eventually the total interaction becomes $\frac{1}{8}(u_{11}-2\sqrt{2}u_{10}) \int dz \rho^2(z)$, which becomes purely attractive if $u_{10}\gg u_c\equiv u_{11}/(2\sqrt{2})$\cite{note}.

\section{Summary}

In summary, we have demonstrated that a multiple-channel scattering resonance(MCSR) can be achieved for spinor bosons confined in 1D geometry. The two-body and many-body properties revealed in this work are expected to be easily probed in current cold atoms experiments.
%how to use TMF achieving strongly interacting spinor bosons in the confined 1D geometry, and revealed the associated many-body properties of this system. The discovered off-M scattering and simultaneous multiple-channel resonances indispensably rely on the interplay of TMF, the confinement and the bare 3D interactions.
%The mechanism for these phenomena is so intrinsic that it can also apply to other confined geometry , such as the confined 2D systems, or high-spin fermions which ic

Finally, we remark that the proposal of MCSR widely applies to other high-spin systems which allow more-than-one collision channels, and other confined geometries such as 2D.
%It is promising that a new pairing superfluidity will emerge for high-spin fermions in 2D near MCSR.
Moreover, rf field can be replaced by any field that allows spin-flips, such as a rotating magnetic field generating spin-orbit couplings\cite{soc}.
%In that case, MCSR is expected to induce even intriguing many-body phenomena due to additional spin-orbit correlations.
Given the wide applicability and special properties of MCSR, we expect this new type of scattering resonance will
%Overall, the new type of scattering resonance as MCSR is expect to
induce a lot more intriguing many-body physics in the subject of spinor systems, such as the ground state structure and topological defects of a 2D BEC, the pairing superfluidity of high-spin fermions, the interplay effect with spin-orbit correlations and so on.
%, and potentially will stimulate more explorations along this route.

The author thanks Randy Hulet, Shuai Chen, Hui Zhai, Tin-Lun Ho, Xi-Wen Guan and Jian Li for stimulating discussions, and Wenbo Fu for early contribution on this project. This work is supported by NSFC under Grant No. 11104158, No. 11374177, and programs of Chinese Academy of Sciences.

\begin{widetext}

\appendix

\section{Derivation of $u-$matrix (Eq.(\ref{u-mtx}))}

First, we decouple the non-interacting two-body Hamiltonian as follows,
\begin{equation}
H^{(0)}_1+H^{(0)}_2=\left(-\frac{\bigtriangledown_{X}^2+\bigtriangledown_{Y}^2+\bigtriangledown_{Z}^2}{4m}+m\omega_{\perp}^2 (X^2+Y^2)\right)+
\left(-\frac{\bigtriangledown_{x}^2+\bigtriangledown_{y}^2+\bigtriangledown_{z}^2}{m}+\frac{m}{4}\omega_{\perp}^2 (x^2+y^2)+\Omega(\sigma_x^1+\sigma_x^2)\right)  \label{H0}
\end{equation}
Here the first bracket represents the Hamiltonian for center-of-mass motion ${\bf R}\equiv(X,Y,Z)=\frac{{\bf r}_1+{\bf r}_2}{2}$ with effective mass $2m$; the second bracket includes the relative motion ${\bf r}\equiv(x,y,z)={\bf r}_1-{\bf r}_2$ with mass $m/2$, and the spin part according to a transverse magnetic field. Since the interaction $U$ is only relevant to the relative motion and their spins, we only consider the second bracket in Eq.(\ref{H0}) when solving two-body problem in the following.

The eigen-state for non-interacting Hamiltonian (second bracket in Eq.(\ref{H0})) can be expressed as $|n_x,n_y,k\rangle|\alpha_1\beta_2\rangle$, where the first part describes the relative motion with wave function
\begin{equation}
\langle x,y,z | n_x,n_y,k\rangle=\phi_{n_x}(x) \phi_{n_x}(y) \frac{e^{ikz}}{\sqrt{L_z}} ,
\end{equation}
and the second part shows the spin configuration with $\alpha,\beta=+$ or $-$, and $|\pm\rangle=(|\uparrow\rangle\pm|\downarrow\rangle)/\sqrt{2}$. The corresponding energy spectrum is
\begin{equation}
E_{n_x,n_y,k;\alpha_1\beta_2}=(n_x+n_y+1)\omega_{\perp} +\frac{k^2}{m}+(\epsilon_{\alpha}+\epsilon_{\beta}),
\end{equation}
here $\epsilon_{\pm}=\pm \Omega$. The threshold scattering energy is therefore $E_{th}=\omega_{\perp}-2\Omega$ when $n_x=n_y=0,\ k=0,\ \alpha=\beta=-$.

%where $\phi_n$ is the eigen-state for 1D harmonic oscillator with index $n$.

Making use of the Lippman-Schwinger equation $U|\Psi\rangle=T|\Psi^{(0)}\rangle$, and the T-matrix $T=U+UG_{0}T$, we get\begin{equation}
U|\Psi\rangle =U|\Psi^{(0)}\rangle+UGU|\Psi\rangle, \label{LS2}
\end{equation}
then combining with Eq.(\ref{U3d}) in the text, we obtain the matrix
equation expanded in $\{|M\rangle\langle N|\}$ spin space, (see also
Eq.(\ref{f3d}) in the text)
\begin{equation}
\Big[\left(
  \begin{array}{ccc}
      &   &   \\
      & U &   \\
      &   &   \\
  \end{array}
\right)^{-1}-\left(
  \begin{array}{ccc}
      &   &   \\
      & \tilde{G}_0\equiv G_0({\bf r=0}) &   \\
      &   &
  \end{array}
\right)\Big]\left(
         \begin{array}{c}
          F_1 \\
           F_{0} \\
           F_{-1} \\
         \end{array}
       \right)=\left(
         \begin{array}{c}
           \tilde{\Psi}^{(0)}_1 \\
           \tilde{\Psi}^{(0)}_0 \\
           \tilde{\Psi}^{(0)}_{-1} \\
         \end{array}
       \right), \label{LS_3d}
\end{equation}
here matrix $(U)=diag(U_1,U_0,U_{-1})$;
%$\left[\tilde{G}_0\right]_{MN}\equiv $
$\left[\tilde{G}_0({\bf r})\right]_{MN}=\langle {\bf r},M|G_0|{\bf
0},N\rangle$; $\tilde{\Psi}^{(0)}_M=\langle M|\Psi^{(0)}({\bf
r=0})\rangle$. With the information of $F_M$ from above matrix
equation, the wf can be deduced straightforwardly, which is
Eq.(\ref{psi3d}) in the text.

For low-energy scattering $E\ll E_{th}+ 2\omega_{\perp}$ and at $z\rightarrow
\infty$, in the full wave function all excited ${\bf n}$(with $n_x+n_y>0$) modes decays away except the
lowest ${\bf n_0} (n_x=n_y=0)$, i.e., the wave function is effectively propagating in 1D as
\begin{eqnarray}
\Psi({\bf r})&\rightarrow &\phi_0(x)\phi_0(y)\Big\{\psi^{(0)}(z)+
g_0(z) (f_{1}|1\rangle+f_{0}|0\rangle+f_{-1}|-1\rangle)\Big\}
\end{eqnarray}
with $f_M=F_M{\phi_0^*}^2(0)$, and $g_0$ is the non-interacting Green function for 1D system (see definition in the text). Alternatively, the wave function along $z$ can be generated
effectively through a 1D interaction
\begin{eqnarray}
\langle z|u |\psi\rangle=\delta(z) (f_{1}|1\rangle+f_{0}|0\rangle+f_{-1}|-1\rangle),\label{U1d}
\end{eqnarray}
together with the Lippman-Schwinger equation
\begin{equation}
\Big[\left(
  \begin{array}{ccc}
      &   &   \\
      & u &   \\
      &   &   \\
  \end{array}
\right)^{-1}-\left(
  \begin{array}{ccc}
      &   &   \\
      & \tilde{g}_0\equiv  g_0(z=0) &   \\
      &   &
  \end{array}
\right)\Big]\left(
         \begin{array}{c}
          f_1 \\
           f_{0} \\
           f_{-1} \\
         \end{array}
       \right)=\left(
         \begin{array}{c}
          \tilde{\psi}^{(0)}_1 \\
          \tilde{\psi}^{(0)}_0 \\
          \tilde{\psi}^{(0)}_{-1} \\
         \end{array}
       \right). \label{LS_1d}
\end{equation}

Compare Eq.(\ref{LS_1d}) with Eq.(\ref{LS_3d}), and recall the
relations that $f_M=F_M{\phi_0^*}^2(0),\
\tilde{\psi}^{(0)}_M=\tilde{\Psi}^{(0)}_M/\phi_0^2(0)$, we obtain
\begin{equation}
\left(
  \begin{array}{ccc}
      &   &   \\
      & u &   \\
      &   &   \\
  \end{array}
\right)^{-1}-\left(
  \begin{array}{ccc}
      &   &   \\
      & \tilde{g}_0&   \\
      &   &
  \end{array}
\right)=\frac{1}{|\phi(0)|^4}\Big[\left(
  \begin{array}{ccc}
      &   &   \\
      & U &   \\
      &   &   \\
  \end{array}
\right)^{-1}-\left(
  \begin{array}{ccc}
      &   &   \\
      & \tilde{G}_0&   \\
      &   &
  \end{array}
\right)\Big],\label{3d_1d}
\end{equation}
which gives rise to Eq.(\ref{u-mtx}) in the text.

Next we show the detailed procedure how to evaluate $\tilde{G}_0^{ex}$ and $u-$matrix in Eq.(8).  To expand $\tilde{G}_0^{ex}$, one needs to insert a complete set of eigen-states $\{ |n_x,n_y,k\rangle|\alpha_1\beta_2\rangle\}$ and sum over all contributions from these energy states. We will see in the following that for a diagonal element of $\tilde{G}_0^{ex}$, the summation will have ultraviolet divergence, which will be compensated by the same divergence in $U^{-1}$. Eventually each element of $u$ is physically a finite value.

As the ultraviolet divergence in energy space corresponds to the short-range singularity of the wave function ($\sim 1/r$) as inter-particle distance $r\rightarrow 0$, in the following we will try to exact the physical value of $\tilde{G}_0^{ex}$ by evaluating the Green function in coordinate space and subtracting $1/r$ singularities. Explicitly, take one diagonal element($M=N=1$) of example, we have
\begin{eqnarray}
\left[G_0^{ex}({\bf r})\right]_{11}&=&\langle {\bf r};M=1| G_0^{ex}
|
{\bf 0};M=1\rangle  \nonumber\\
&=&\sum_{n_1+n_2>0}^{\infty}\phi_{n_1}(x)\phi_{n_1}^*(0)\phi_{n_2}(y)\phi_{n_2}^*(0)\int_{-\infty}^{\infty}\frac{d
k}{2\pi}e^{ikz}
\left( \sum_{\alpha,\beta} \frac{\langle M=1|\alpha_1\beta_2\rangle\langle \alpha_1\beta_2|M=1\rangle}{\Delta
E-(n_1+n_2)\omega_{\perp}-\frac{k^2}{m}-(\epsilon_{\alpha} + \epsilon_{\beta}+2\Omega)+i\delta} \right)
\label{eq1}
\end{eqnarray}
with $\Delta E=E-E_{th}$; $\langle M=1|\alpha_1\beta_2\rangle=\xi_{\alpha}^{\uparrow}\xi_{\beta}^{\uparrow}$ and $\xi_{\alpha}^{\uparrow}=\xi_{\beta}^{\uparrow}=1/\sqrt{2}$. Note that in order for the non-zero value of above equation as ${\bf r}\rightarrow 0$, $n_x+n_y$ should be an even integer ($=2,4,...$), therefore as long as $\Delta E<2\omega_{\perp}$ the denominator inside the bracket is always negative. For this low-energy scattering ($\Delta E\ll 2\omega_{\perp}$), Eq.(\ref{eq1}) is transformed to
\begin{eqnarray}
-\sum_{\alpha,\beta}  |\langle M=1|\alpha_1\beta_2\rangle|^2  \int_0^{\infty} dt A(t) e^{(\Delta E) t} e^{-(\epsilon_{\alpha} + \epsilon_{\beta}+2\Omega) t}  \int_{-\infty}^{\infty}\frac{d k}{2\pi}e^{ikz} e^{-\frac{k^2}{m} t}, \label{eq2}
\end{eqnarray}
and $A(t)$ can be obtained by making use of the propagator of 1D harmonic oscillator,
\begin{eqnarray}
A(t)&=&(\sum_{n_1=0}^{\infty}e^{-n_1\omega_{\perp}
t}\phi_{n_1}(x)\phi_{n_1}^{*}(0))(\sum_{n_2=0}^{\infty}e^{-n_2\omega_{\perp}
t}\phi_{n_2}(y)\phi_{n_2}^{*}(0))
-\phi_{0}(x)\phi_0^*(0)\phi_{0}(y)\phi_0^*(0)\nonumber\\
&=&\frac{1}{\pi a_{\perp}^2}(\frac{1}{1-e^{-2\omega_{\perp}
t}}e^{-\frac{x^2+y^2}{2a_{\perp}^2}\coth(\omega_{\perp}
t)}-e^{-\frac{x^2+y^2}{2a_{\perp}^2}}).\ \ \ \ \ \ (a_{\perp}=\sqrt{\frac{2}{m\omega_{\perp}}})
\end{eqnarray}

Further using the identity
\begin{eqnarray}
\int_{-\infty}^{\infty}\frac{d k}{2\pi}e^{ikz}
e^{-\frac{k^2}{m}t} = \frac{1}{a_{\perp}\sqrt{2\pi
\tau}}e^{-\frac{z^2}{2a_{\perp}^2\tau}},\ \ \ \ \ \ \ \ (\tau=\omega_{\perp} t)
\end{eqnarray}
Eq.(\ref{eq2}) is further reduced to
\begin{eqnarray}
&&-\sum_{\alpha,\beta}  |\langle M=1|\alpha_1\beta_2\rangle|^2
\frac{m}{(2\pi)^{3/2} a_{\perp}}\int_0^{\infty} \frac{d\tau}{\sqrt{\tau}} e^{\frac{\Delta
E}{\omega_{\perp}} \tau} e^{-\frac{\epsilon_{\alpha} + \epsilon_{\beta}+2\Omega}{\omega_{\perp}} \tau}
(\frac{1}{1-e^{-2\tau}}e^{-\frac{x^2+y^2}{2a_{\perp}^2}\coth\tau}-e^{-\frac{x^2+y^2}{2a_{\perp}^2}})
e^{-\frac{z^2}{2a_0^2\tau}}.  \label{eq3}
\end{eqnarray}
It is then noticed this integral has singularity at $\tau\rightarrow 0$ and as $r=\sqrt{x^2+y^2+z^2}\rightarrow 0$, and the singularity is given by
\begin{equation}
-\frac{m}{(2\pi)^{3/2} a_{\perp}} \int_{0}^{\infty}\frac{d\tau}{2\tau^{\frac{3}{2}}}e^{-\frac{r^2}{2a_{\perp}^2\tau}}=-\frac{m}{4\pi r}.
\end{equation}
Apart from the singularity term, the (physical) constant term in the asymptotic form
\begin{equation}
\lim_{r\rightarrow 0} G_0^{ex}({\bf r})_{11}= -\frac{m}{4\pi r} +C_{11} \label{asym}
\end{equation}
can be extracted as
\begin{eqnarray}
C_{11}&=&- \sum_{\alpha,\beta}  |\langle M=1|\alpha_1\beta_2\rangle|^2
\frac{m}{(2\pi)^{3/2} a_{\perp}}\int_0^{\infty} \frac{d\tau}{\sqrt{\tau}} \left( e^{\frac{\Delta
E}{\omega_{\perp}} \tau} e^{-\frac{\epsilon_{\alpha} + \epsilon_{\beta}+2\Omega}{\omega_{\perp}} \tau}
(\frac{1}{1-e^{-2\tau}}-1)-\frac{1}{2\tau} \right).   \label{C11}
\end{eqnarray}
Recalling that $\frac{1}{U_{11}}= \frac{m}{4\pi a_1} -\frac{1}{V}\sum_{\bf k} \frac{1}{(k^2/m)}=\frac{m}{4\pi a_1} -\lim_{r\rightarrow 0}  \frac{m}{4\pi r}$, and combining with Eq.(\ref{asym}), we get the element
\begin{equation}
\big[ U^{-1}-\tilde{G}_0^{ex}\big]_{11}=\frac{m}{4\pi a_1} - C_{11}.
\end{equation}

Similarly one can obtain diagonal $C_{MM}$ for $M=0,-1$, by
replacing $M=1$ with $M=0,-1$ in Eq.(\ref{C11}). The one-dimensional
integration in term of imaginary time $\tau$ can be performed
straightforwardly by numerics. In the case of $\Omega=0$, we have
$C_{MM}=\frac{m}{4\pi a_{\perp}}c_0$ that are identical for all $M$,
with $c_0=1.46$ obtained previously for conventional CIR(see
Ref.[2]).

The same strategy can be used to calculate off-diagonal elements of matrix $\left(U^{-1}-\tilde{G}_0^{ex}\right)$, and one can find that the divergence of these elements at short distance will be absent. Explicitly we have $(M\neq N)$
\begin{equation}
C_{MN}=- \sum_{\alpha,\beta}  \langle M|\alpha_1\beta_2\rangle \langle \alpha_1\beta_2 | N\rangle
\frac{m}{(2\pi)^{3/2} a_{\perp}}\int_0^{\infty} \frac{d\tau}{\sqrt{\tau}}  e^{\frac{\Delta
E}{\omega_{\perp}} \tau} e^{-\frac{\epsilon_{\alpha} + \epsilon_{\beta}+2\Omega}{\omega_{\perp}} \tau}
(\frac{1}{1-e^{-2\tau}}-1); \label{Cmn}
\end{equation}
\begin{equation}
\big[ U^{-1}-\tilde{G}_0^{ex}\big]_{MN}= - C_{MN}
\end{equation}

Knowing all the elements of matrix
$\left(U^{-1}-\tilde{G}_0^{ex}\right)$, the $u-$matrix can be
obtained through Eq.(\ref{u-mtx}).

\end{widetext}


\begin{thebibliography}{}
\bibitem{footnote_dif} The "mixed-channel scattering" in Ref.\cite{CC} and \cite{QDF} refers to scattering between different orbital angular momenta, which is due to the interplay effect of short-range inter-particle interaction, the magnetic dipole interaction, the fine and hyperfine coupling, and the magnetic Zeeman energy. While the "multiple-channel" in our work refers to scattering between different total spin magnetizations, due to the interplay of the s-wave interaction, the confinement and the spin-flipping field. As a result, their physical consequences are also substantially different.
\bibitem{footnote_rf}To avoid the system being polarized by rf field, we consider $\Omega\ll E_F$, where $E_F$ is the Fermi energy of identical fermions with the same density as bosons.
\bibitem{note}% In order for $u_{10}$ and $u_{11}$ to satify this relation, we have checked from Fig.2 that one has to fine-tune the parameters very close to the location of MCSR. Therefore to turn the system to purely attractive is in practice hard to achieve. The most practical situation to realize in experiment is the coexisting of attraction(for $\downarrow$) and repulsion($\uparrow$), as shown by Fig.3(a3,b3).
The ratio of $u_{10}$ to $u_{11}$ near MCSR is directly given by the
ratio of their individual width, which is tunable through the rf
strength and interaction parameters. For a weak rf field, this ratio cannot be too large, so the system will hardly be fully polarized or turn purely attractive.
\end{thebibliography}
\end{document}